# Power-efficient Routing & Increased Yield Approach For WSNs

Kalpana Sharma, Anurag S. Rathor, S. R. Biradar, M.K Ghose
Department of Computer Engineering,
Sikkim Manipal Institute of Technology
Majitar, Rangpo, East Sikkim -737 132, SIKKIM

*Abstract—* **The sensor nodes in a Wireless Sensor Network are generally constrained with limited power supply. Efficient power management is a must for any sensor network to keep the sensor nodes in the network to be operational for a longer period of time this increasing the lifetime of the sensor network. Hierarchy based routing enables the sensor networks to be deployed in larger areas. In this paper we present a hierarchical cluster based routing protocol which improves the scalability as the data travels from one cluster level to another covering a greater amount of distance and increases the lifetime of the wireless sensor network by distributing the power dissipation load evenly among all the sensor nodes within the network. Also the time delay in case of critical data to be received by the Base Station has also been lowered.**

## I. Introduction

Wireless Sensor Networks (WSNs) are of greater importance and find their applications in various military surveillances, disaster management and security systems. The sensor nodes are unattended and are energy constrained as they have a limited source of power and die when their power is dissipated completely. These sensors are expected to work until their power is fully utilized. Thus efficient network and power management is crucial for any wireless sensor network in order to maximize their lifetime.

The sensors are capable of sensing the data from the environment in which they are deployed, processes that data and transmit it to the base-station (BS). The sensor circuit senses the environment and converts the signals into electrical signals which are then transmitted to the BS using a transmitter via a routing node [5]. The transmitter circuit uses more amount or energy as compared to the receiver circuit. The power dissipated in transmission and receiving is calculated with the help of the following equations [2].

*Transmitting*:
$E_{Tx}(k, d) = E_{Tx\text{-elec}}(k) + E_{Tx\text{-amp}}(k, d)$
$E_{Tx}(k, d) = E_{elec} * K + \epsilon_{amp} * k * d^2$

*Receiving*:
$E_{Rx}(k) = E_{Rx\text{-elec}}(k)$
$E_{Rx}(k) = E_{elec} * k$

## HIERARCHICAL CLUSTERING

The hierarchical cluster based routing is advantageous as the scalability and the power efficiency in the sensor network improves [7]. In this hierarchical based architecture, nodes with higher power levels perform the fusion of data gathered from the other sensor nodes and transmit the aggregated data to the base-station (BS) while the nodes with low power levels only perform the sensing of the environment. They transmit the sensed data to the higher node, known as the cluster-heads (CHs) which are at a lesser distance than the base station. The cluster formation and the assignment of special tasks to the cluster heads (CHs) reduce the power dissipation within a particular cluster, which improves the scalability of the sensor network. Also by aggregating the sensed data, the amount of data to be transmitted to the base-station (BS) is reduced and the lifetime of the overall sensor network is increased. As the data travels from a lower cluster level to a higher cluster level, it covers more distance and the data travels faster to the base-station (BS).

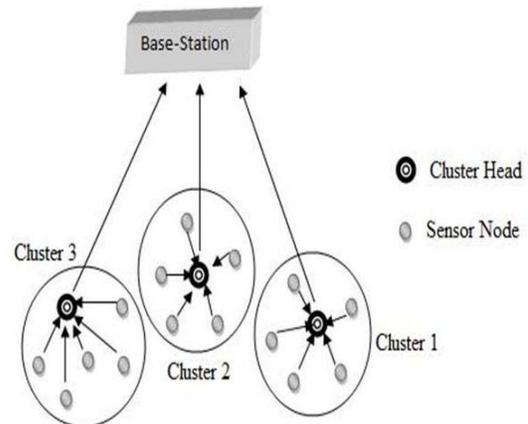

Fig. 1

Figure 1 shows a hierarchical cluster based routing setup in which the sensor nodes transmit the sensed data to the cluster head which in turn transmits the aggregated data to the BS.

## II. SURVEY AND RELATED WORK

We have studied few existing network routing protocols which are capable of routing the aggregated data in an energy efficient manner [8]. They are briefly discussed below and compared in a tabular form in table 1.

LEACH
Low Energy Adaptive Clustering Hierarchy (LEACH) is a hierarchical cluster based routing protocol proposed by Heinzelman et. al. [1]. This protocol incorporates the formation of clusters and cluster heads (CHs) for the respective clusters in which all the other sensor nodes send the data to the cluster head (CH). The received data is then aggregated and is sent to the base-station (BS) periodically by the cluster head which reduces the amount of data that is to be transmitted to the base-station. The role of the cluster head (CH) is rotated among the other sensor nodes in the cluster so as to evenly distribute the power load between the sensor nodes in a particular cluster. A TDMA/CDMA MAC is used for avoiding the collisions among the clusters and within the clusters.

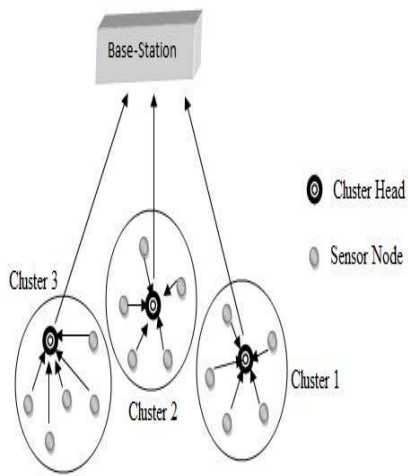

Fig 2

*Working Principle:* The LEACH protocol functions in two different phases. The *setup phase* and the *steady state phase* [8]. The formation of clusters and selection of the cluster heads is done during the setup phase and the aggregated data is transmitted to the base-station during the steady state phase which is of greater duration than the setup phase. During the setup phase, a random number $r$, between 0 and 1, is selected by the sensor nodes. If this random number is less than a threshold value $T(n)$, that sensor node is selected as the cluster head. The threshold value $T(n)$ is calculated as follows [8]:

$T(n) = p / [1-p(r \bmod(1/p))]$    if n ∈ G

Where, p is the predetermined number of sensor nodes, r is the random number and G is the set of nodes that are involved in the CH selection that have not been selected as cluster heads in the last $(1/p)$ round. After the selection, the cluster heads sends an advertisement to all the other sensor nodes in the network. The formation of clusters is based upon the signal strength of this advertisement. After the cluster formation, a TDMA schedule is created assigning time slots to the sensor nodes for data transmission. After the cluster formation and the selection of the cluster heads, the network goes into steady state phase where the aggregated data from the sensor nodes is sent to the base-station by the cluster heads. The network again goes back into the setup phase after a predetermined time period to select a new set of cluster heads as to rotate the role of the cluster heads among the nodes of a cluster.

The network lifetime is increased as the load of power dissipation is evenly distributed among the nodes in the cluster. Also the amount of data to be transmitted is less which in turn reduces the latency of the network.

The LEACH protocol is not suitable for networks deployed in large areas. Also the predetermined cluster heads may not be uniformly distributed. The path taken by the aggregated data to reach the base station is not optimal.

PEGASIS
The Power-Efficient Gathering in Sensor Information Systems (PEGASIS) proposed in [2] is an improvement over the LEACH protocol. It is a near optimal chain-based protocol. The idea of cluster formation and cluster head is discarded in PEGASIS. Instead of multiple nodes, a single node in the chain communicates with the base-station. The sensor nodes in this protocol only communicate with a single node closest to them and communication with the base-station is done in rounds so that the power dissipation in communicating with the base-station is distributed evenly among all the nodes.

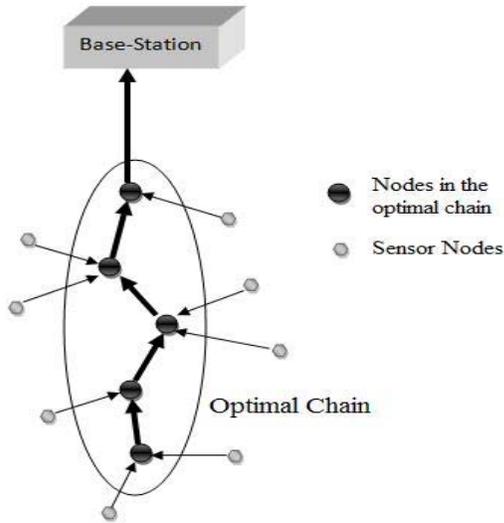

Fig 3

*Working Principle:* PEGASIS assumes that all the sensor nodes maintain a database of the location of all the other nodes in the network. Each node determines the distance of its neighboring nodes using the signal strength and adjusts the signal strength only to communicate with the closest node. In PEGASIS, the sensor nodes closest to each other are in the chain and they form a path to transmit the aggregated data to the base-station. The chain is constructed using Greedy algorithms. Each sensor node sends the sensed data to its closest node in the chain. The data is aggregated at each node in the chain and finally only the aggregated data is sent to the base-station. The lifetime of each node is increased as they only have to communicate with their closest node which, as a result increases the network lifetime.

Delay is caused in data transmission from the distant node in the chain. There is significant overhead as the nodes need the know-how about the other node location and the path for transmitting data.

To overcome the problem of delay occurrence in transmitting the aggregated data to the base-station (BS) an extension to PEGASIS, called Hierarchical-PEGASIS was introduced in which the transmission of the data was allowed only by the spatially separated sensor nodes. This ensured parallel data transmission and reduced the delay.

TEEN
Threshold-sensitive Energy Efficient sensor Network (TEEN) is a hierarchy based routing protocols proposed in [3], for time-critical applications. The region is sensed continuously by the sensor nodes but the sensed data is transmitted less frequently. The cluster heads (CHs) broadcasts a hard threshold, which is the threshold value of the sensed data and a soft threshold, which is a small change in the hard threshold value of the sensed data to all the other sensor nodes in a cluster. The soft threshold instigates the sensor node to switch on its transmitter and transmit the data.

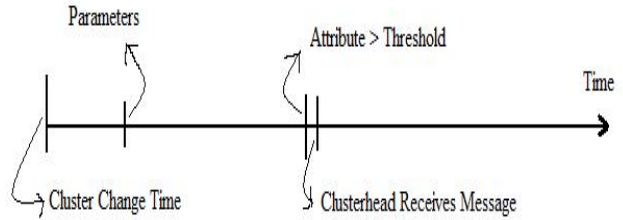

Fig 4

*Working Principle:* In TEEN, a hard threshold value and a soft threshold value is sent to all the other sensor nodes by their respective cluster heads (CHs). The sensor nodes begin to transmit data by switching on their transmitters when they sense a change in the soft threshold value. Transmission of data occurs only when the sensed data is in the range of interest of the user.

*Adv/Disadv:* TEEN protocol reduces the number of transmissions by only transmitting the data only when the sensed data is of interest to the user.

The major disadvantage of the TEEN protocol is that, if the threshold values are not received, the sensor nodes will not communicate and the user will not receive any data either.

APTEEN
*Def:* Adaptive Periodic Threshold-sensitive Energy Efficient sensor Network (APTEEN), is a hybrid protocol which was proposed in [4], is also for time-critical applications. In APTEEN, according to the user needs and the application type, the threshold values used in TEEN are changed at some specific time intervals.

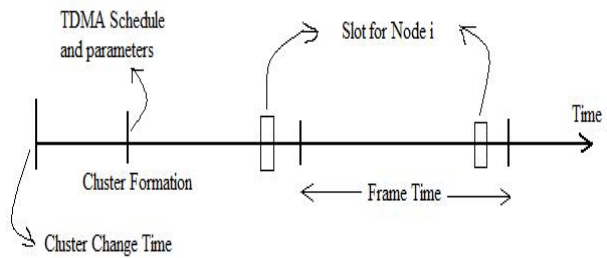

Fig 5

*Working Principle:* In APTEEN, few parameters such as the Attributes (A), Hard Threshold (HT), Soft threshold (ST), Schedule and Count Time (CT) are sent to the other sensor nodes in the cluster by the respective cluster heads (CHs). When the sensed data value is greater than the HT, the data is transmitted only when there is a change in that value. Each sensor node in the cluster is given a time slot using a modified TDMA schedule for transmission. When a sensor node does not transmit data for a time period equal to the CT, it is forced to sense again and retransmit the data.

*APTEEN* is flexible, as the power consumption is controlled by the user by changing the count time (CT) and the threshold values.

The implementation of the threshold values and the count time (CT) is complex. Also the overhead increases.

### III. COMPARATIVE STUDY

We compared a few hierarchical cluster based routing protocols and summarized them in the below given tabular form. Here it must be noted that for our literature study we considered the cluster based routing protocols including the PEGASIS. But for the comparison of our results we have excluded PEGASIS as it does not facilitate the concept of formation of clusters. We compared the results of our protocol generated with that of LEACH, APTEEN and TEEN. The comparison between the routing protocols was based on many metrics as given below.

| Protocols / Metrics | LEACH | PEGASIS | TEEN & APTEEN |
|---|---|---|---|
| Optimal Path | No | Yes | No |
| Position Known | No | Yes | No |
| Power Usage | Max | Max | Max |
| Scalability | Good | Good | Good |
| Network Lifetime | V Good | V Good | V Good |
| Data Fusion | Yes | No | Yes |
| State Complexity | CHs | Low | CHs |

Table 1

### IV. PROPOSED WORK

We propose a hierarchical cluster based routing protocol which minimizes the energy dissipation caused due to long transmission and large data aggregation. Rather than a single cluster head (CH) performing both data aggregation and transmitting it to the base-station, we have divided this load among two sensor nodes which acts as the *Data Cluster Head (DCH)* and *Routing Cluster Head (RCH)* in a cluster. In this protocol, the sensor nodes are relieved from the load of calculations that are required from the cluster head selection as in [1]. Instead the base station randomly chooses few nodes based upon proximity such that the whole of the deployed region is covered and is divided in clusters. We have assumed that the sensor nodes are deployed almost uniformly over the region. After the formation of the clusters, the base-station (BS) selects two sensor nodes in each cluster as the *Data Cluster Head (DCH)* and the *Routing Cluster Head (RCH)*. The node preferably in the centre of each cluster is selected as the DCH and the node with the minimum distance to the base-station in that cluster is selected as the RCH. Among these routing cluster heads (RCHs), the RCH that is closes to the base-station is selected as the *Principle Cluster Head (PCH)* which will receive the aggregated data from all the other cluster heads (CHs) and transmit it to the base-station (BS). The PCH also acts as the RCH for the cluster in which it belongs. After that, these cluster heads (CHs) broadcast a message to all the other sensor nodes in their respective clusters that they are the CHs and the data will be routed to the base-station via them. This reduces the power dissipation of the sensor nodes as they only need to communicate with the local sensor nodes in their respective clusters. Once these cluster heads are selected, the sensor nodes are then provided with a TDMA time schedule for the transmission of data.

The sensor nodes are then active and start sensing the environment. The sensor nodes have a prior knowledge of the desired range of data. If the sensed data lies in the range then the sensor nodes transmits the data to the DCH which aggregates the data received from all the other sensor nodes and transmits it to the RCH. The RCHs of all the clusters transmit the aggregated data to the principle cluster head (PCH) which in turn, transmits the data to the base-station (BS). If the value of the aggregated data is more than the maximum value of the range, the RCH transmits the data directly to the base-station as that data may be critical and delay in these kinds of data is not desirable. If the value of the sensed is less than the minimum value of the range, the sensor nodes do not transmit, they keep their transmitter switched of which saves a considerable amount of power. After a predetermined level of the cluster-heads energy is spent, the role of the different CHs is rotated and a new set of DCHs, RCHs and PCH nodes will be selected. The new CHs will be selected depending upon the signal strength between the peer nodes. The current CHs will select the nodes closest to them in a cluster and they will be the new CHs. For this we have assumed that the sensor nodes maintain a database about their distance from the other sensor nodes to which they are communicating. They are not required to have the knowledge of the distance of every node in the network as in [2]. By rotating the role of the CHs, the load of power dissipation is distributed evenly among all the sensor nodes in a cluster which accounts in increasing the lifetime of the overall sensor network. Performing data aggregation at the cluster heads minimizes the number of transmission from one cluster level to the higher cluster level. This also minimizes the number of transmissions to the base-station, as a result, low power dissipation and increased network lifetime.

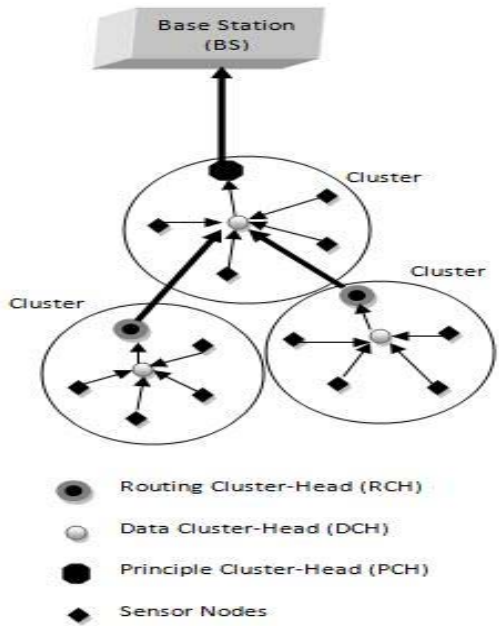

Fig 6

## V. PERFORMANCE METRICS

To evaluate our routing protocol we used a few metrics to compare it with the other existing protocols. The metrics used are:

- *Power consumption in transmission* – Provides the amount of energy dissipated by a sensor node to transmit the data to the next higher level node. Lower the power consumption better is the network yield.
- *Lifetime of a node* – This gives the time for which a sensor node is operational.
- *Average lifetime of nodes* – Provides a good overview of the lifetime of the sensor network.
- *Time taken in cluster formation* – The time until which the first data packet is received by the cluster head.
- *Average Delay per packet* – The average time a data packet takes to reach the cluster head.
- *Lifetime of the network* – Time after which the first node is not operational i.e. dead.
- *Network yield* – Calculated as the total number of data packets received at the cluster head divided by the time taken for simulation.

## VI. EXPERIMENTAL SETUP

To get the performance results of our routing protocol we used the OMNeT++ v3.2 simulation tool with the mobility framework.

Our experimental setup for performing the simulation of the proposed routing protocol had 100 sensor nodes deployed randomly over a 100X100 meter square area. The bandwidth of the link between the peer nodes was 10 Kbits/s. The cost of transmission and receiving was calculated by the formulas mentioned above in the section I. The minimum amount of energy required for a node to be a CH was 35% of the total energy and the amount for a sensor node to be operational was 5% of the total energy. The desired range was specified as 30 -60 degree centigrade. The temperature above 60 degrees was considered as critical data and the temperature below 30 degrees was considered as undesired data during which the sensor nodes went into sleep mode turning off their receivers.

## VII. RESULTS & DISCUSSIONS

We simulated our protocol and compared the generated data with that of the existing routing protocols. Note that, here we did not compare the result with that of PEGASIS. For our comparison we only considered the LEACH, APTEEN and TEEN routing protocols as they are *Hierarchical Clustering* based routing protocols.

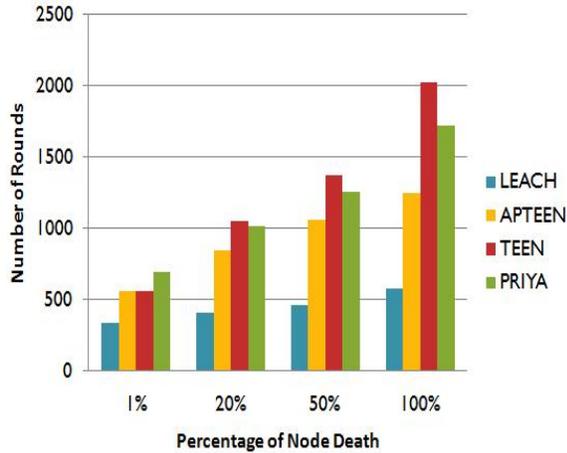

Fig 7

The above figure shows the graph of percentage of node deaths v/s number of rounds. Here we can observe that our protocol outperforms the other protocols except for TEEN in which the data is sent periodically, whereas in case of PRIYA, the data is sent continuously. The transmission of data takes place only when the sensed data is within the desired range. Also, the rotation of the role of CHs was determined on the basis of spent energy. While in case of the other protocols, the CH role was rotated after a predetermined period of time.

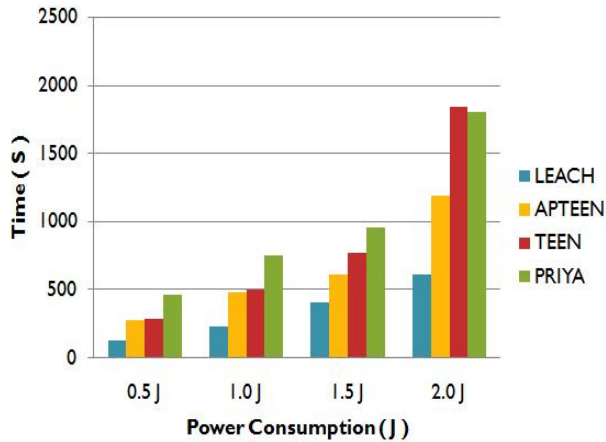

Fig 8

Figure 8 shows the power consumptions. At the initial stages, PRIYA outperforms the other protocols but in the later stages the power consumptions increases. As the node deaths increases, the remaining nodes need to have a longer communication with the CHs and the BS as compared to the earlier stages.

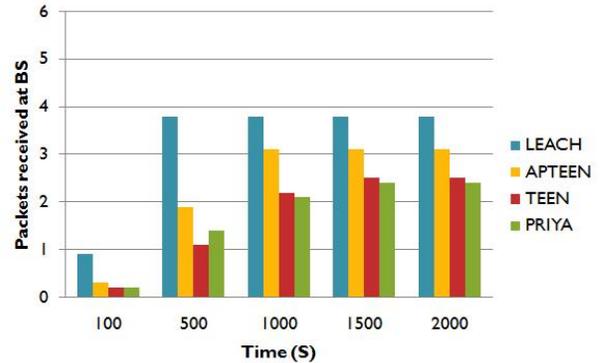

Fig 9

Figure 9 depicts the number of data packets received by the base-station. The lesser the number of received data packets, the lesser communication required, which decreases the power consumptions. Also, less data packets means, more accurate data received.

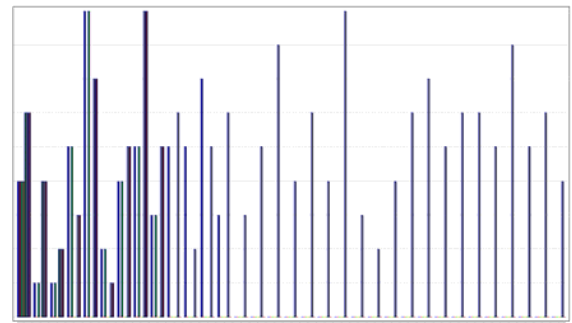

Fig 10

The above figure was generated by the simulator, which depicts the power dissipation of the nodes. The higher bars represent the nodes communicating directly with the BS in case of critical data. The lowest bars represent the nodes sensing redundant data, which requires no transmission.

## VIII. CONCLUSION

In this paper we propose PRIYA (Power-efficient Routing & Increased Yield Approach), a hierarchical cluster based routing protocol which is ideal for sensor networks deployed in large regions when delay caused in the transmission of critical data is an important factor. Our protocol facilitates the concept of Hierarchical Cluster-Based routing for efficient routing of data and the sensor nodes communicate directly with the BS minimizing the delay occurred in transmitting critical data. The evenly distribution of work load among the nodes ensure lesser power dissipation hence increasing the yield. The protocol is designed to fit for a particular environment by allowing the user to define the range of desired data resulting in greater and efficient yield along with power efficient routing of data.

## IX. REFRENCES